\documentclass[sigconf,authorversion,nonacm]{acmart}
\DeclareMathOperator*{\argmin}{argmin}
\usepackage[ruled,vlined,linesnumbered]{algorithm2e}
\usepackage{gensymb}
\usepackage{graphicx}
\usepackage{subcaption}
\usepackage{mwe}
\usepackage{enumitem}
\usepackage{caption} 
\usepackage{adjustbox}
\usepackage{bm}
\AtBeginDocument{%
  \providecommand\BibTeX{{%
    \normalfont B\kern-0.5em{\scshape i\kern-0.25em b}\kern-0.8em\TeX}}}

\setcopyright{none}

\begin{document}

\title{A Data-Efficient Approach to Behind-the-Meter Solar Generation Disaggregation}
\author{Xinlei Chen}
\affiliation{%
  \institution{University of Alberta}
  \city{Edmonton}
  \country{Canada}}
\email{xinlei1@ualberta.ca}

\author{Moosa Moghimi Haji}
\affiliation{%
  \institution{University of Alberta}
  \city{Edmonton}
  \country{Canada}}
\email{moghimih@ualberta.ca}

\author{Omid Ardakanian}
\affiliation{%
  \institution{University of Alberta}
  \city{Edmonton}
  \country{Canada}}
\email{ardakanian@ualberta.ca}

\renewcommand{\shortauthors}{Chen, et al.}

\begin{abstract}
With the emergence of cost effective battery storage and 
the decline in the solar photovoltaic (PV) levelized cost of energy (LCOE),
the number of behind-the-meter solar PV systems is expected to increase steadily.
The ability to estimate solar generation from these latent systems 
is crucial for a range of applications, including distribution system planning and operation, 
demand response, and non-intrusive load monitoring (NILM).
This paper investigates the problem of disaggregating solar generation from smart meter data
when historical disaggregated data from the target home is unavailable, 
and deployment characteristics of the PV system are unknown.
The proposed approach entails inferring the physical characteristics from smart meter data
and disaggregating solar generation using an iterative algorithm.
This algorithm takes advantage of solar generation data (aka proxy measurements) from a few sites
that are located in the same area as the target home, 
and solar generation data synthesized using a physical PV model.
We evaluate our methods with 4 different proxy settings 
on around 160 homes in the United States and Australia,
and show that the solar disaggregation accuracy is improved
by 32.31\% and 15.66\% over two state-of-the-art methods 
using only one real proxy along with three synthetic proxies.
Furthermore, we demonstrate that using the disaggregated home load rather than the net load data
could improve the overall accuracy of three popular NILM methods by at least 22\%.
\end{abstract}




\maketitle

\section{Introduction}
Solar photovoltaic (PV) generation is the fastest growing renewable energy source today~\cite{iea2020}.
Almost half of this growth is projected to be behind-the-meter (BTM) installations,
which are typically PV systems on the roof of homes and buildings~\cite{dehoog2020}.
High penetration of PV systems introduces new challenges 
for planning and operation of power distribution networks,
requiring the system operators and electric utilities 
to develop low-cost techniques for forecasting and monitoring the solar power injected into their feeders.
These techniques should rely on the data that is commonly available to the utilities,
in particular weather information and net meter data (i.e., home load minus solar generation) 
from advanced metering infrastructure 
with time resolutions in the range of one minute to one hour~\cite{solar_disaggregation_review}.

Several methods are proposed in the literature 
to estimate the solar power generated by BTM PV systems.
They can be divided into three major categories:
\begin{enumerate}[label=(\alph*),topsep=0pt]
\item Methods that rely on satellite and aerial imagery~\cite{dehoog2020} 
to identify PV systems and estimate their physical characteristics, e.g., size, tilt, and orientation.
These approaches provide a rough estimate of the peak production capacity, 
but cannot accurately estimate solar generation at a given time.
\item Methods that rely on a few separately metered solar sites in a geographical area 
to estimate the total solar generation in that area~\cite{Shaker2015, Shaker2016}.
These methods require the knowledge of the total installed capacity of PV systems in the area, 
which is not available in many cases.
\item Methods that apply signal separation techniques to disaggregate solar generation 
from feeder-level measurement or smart meter data \cite{Chen2017, Cheung2018}.
The main advantage of the \emph{solar disaggregation} methods that only require smart meter data 
is that they are widely applicable as they do not require additional information.
\end{enumerate}

Solar disaggregation is closely related to non-intrusive load monitoring (NILM)~\cite{NALM},
the problem of separating a household's total energy consumption, 
measured by a single meter, into individual appliance consumption data.
Despite the apparent analogy between the two problems, 
NILM methods cannot be directly applied to separate solar generation from home load.
This is partly because most NILM methods assume that the operation of each appliance 
can be divided into a finite number of operating states (e.g., ON, OFF, standby), 
and that the power consumption in each state is known and constant.
Solar generation however changes continuously depending on several factors, 
such as solar radiation and temperature,
and can have abrupt changes due to the effect of passing clouds.
This diminishes the efficacy of NILM methods~\cite{Dinesh2017}.
Furthermore, solar generation is subtracted from the home load, 
whereas appliance loads are summed to construct the home load.
This `negative' load confuses many NILM methods.
Thus, it is necessary to develop reliable solar disaggregation methods 
to decompose the net load data into solar generation and home load.
The NILM methods can then be applied to the latter component 
to identify the energy consumption of different appliances.

In this paper we propose a solar disaggregation method 
to accurately estimate the output of a BTM PV system in an offline fashion.
Our method has two key advantages over prior work on solar disaggregation.
First, it requires active power measurements with low temporal resolution, 
which is already collected by ordinary smart meters,
and \emph{proxy measurements} from only one or a few PV systems 
located in the same geographical area.
This is motivated by the strong correlation
between the outputs of PV systems located in the same city
even if their deployment characteristics are not identical.
Utilities often have access to direct solar measurements from several sites in a city,
which can provide proxy measurements for solar disaggregation.
Second, our method has a low computational overhead 
compared to the optimization-based methods proposed in the literature,
making it suitable for large scale implementation.
Our contribution is threefold:
\begin{itemize}
    \item We propose a solar disaggregation method that merely relies on net load data and 
    just a small number of separately metered solar PV systems in the same geographical area.
    We show that real proxy measurements can be replaced with synthetic ones 
    to achieve comparable performance as long as there is one real proxy.
    \item We compare our method against other state-of-the-art methods 
    using two publicly available datasets.
    We find that the proposed method outperforms the baselines in terms of disaggregation accuracy.
    \item We examine how the improved accuracy of solar disaggregation affects 
    the accuracy of three baseline NILM methods, 
    namely Factorial Hidden Markov Model (FHMM)~\cite{15RayNILM}, Sequence-to-Point (Seq2Point)~\cite{Seq2Point}, 
    and Denoising Autoencoder (DAE) \cite{NeuralNILM}.
\end{itemize}
To our knowledge, there is no prior work that quantifies the improvement in the accuracy of NILM techniques
when they run on the disaggregated home load rather than the net load measured by a smart meter.
Our results confirm that performing solar disaggregation prior to NILM offers clear benefits.
\section{Related Work}
\subsection{Solar Disaggregation}
Solar disaggregation methods can be categorized based on the type of models they use for estimating solar power.
In the first category, a physics-based model is used to estimate solar generation.
The model parameters are typically inferred from the available data.
Reference~\cite{Chen2017} uses the PV system's location and net load data 
to build a model for maximum clear sky solar generation.
It then estimates the true solar generation using a general weather model
to account for the percentage reduction in the clear sky solar irradiance 
due to clouds, humidity, precipitation, etc. 
The method requires net load data from two days with clear sky and low energy consumption
that have a large temperature difference.
A solar disaggregation toolkit based on~\cite{Chen2017} is developed in~\cite{bashir2019}, 
which is used as a baseline in our work.


Reference~\cite{Kabir2019} proposes an unsupervised solar disaggregation framework
that does not require net load data or the location of the PV system (i.e., its latitude and longitude).
Instead, the authors estimate the home load by using a physical PV model together with Hidden Markov Model Regression.
They start with an initial guess of the physical PV model parameters and 
iteratively estimate the home load and solar generation from the net load data.
This method can achieve satisfactory performance after many iterations,
which makes it too slow for real-world applications. 
In addition, accurate weather and irradiance data from the vicinity of the the target home 
are necessary for accurate disaggregation.
We use this method as a baseline.
The authors extend this work in~\cite{Kabir2021} to jointly disaggregate the net load data of 
a group of customers into aggregated solar generation and aggregated home load.

In the second category of solar disaggregation methods,
a data-driven black-box model is used for the solar generation.
Some of the methods rely on the aggregate data at
the feeder, distribution or substation transformer level~\cite{Tabone2018, Kara2018, game2020}.
The common shortcoming of these methods is that they are not accurate enough for disaggregating a single home's net load.
A promising method that disaggregates solar generation in the frequency domain is proposed in~\cite{Sossan2018}.
This method is also more suitable for feeder-level disaggregation
because high resolution data are not commonly available at the customer level.

There are many methods that rely on customer-level net meter data.
Reference~\cite{Cheung2018} uses a mixture model of representative customers without PV systems 
to model demands of customers with PV systems.
The authors extend this work in~\cite{Cheung2020} by considering BTM battery storage.
In recent work~\cite{arxiv2020}, a disaggregation method is proposed 
based on the observation that there is a strong correlation between monthly 
nocturnal and diurnal home loads.
Leveraging this, the authors estimate the customer's monthly solar generation and subsequently their hourly solar generation.
A shortcoming of this method is that
each customer has some unique characteristics that cannot be captured 
using the home loads of other customers. 
This could lower the accuracy of this disaggregation method.
On the plus side, it does not require any historical home load or solar generation data from the target home.
It relies on low resolution smart meter data and a number of solar proxies.
In this respect, it is similar to our disaggregation method.
However, the performance of this method is only tested when several solar proxies are available.
In contrast, our method can work with only one solar proxy, 
thanks to additional synthetic proxies that we take advantage of.
Despite the novelty of this approach, 
we cannot use it as a baseline because the proposed method requires 3 years of data, 
which is not included in the two datasets we used.
Besides, the publicly available version of the dataset used in that paper 
only contains the aggregated household demand at secondary distribution transformers.

\subsection{NILM}
Since the early days of smart meter rollouts, NILM has been extensively studied in the literature.
At a high level, NILM methods aim to disaggregate the net load measured at a single point of measurement.
A wide range of sampling rates are explored in the literature, from less than 1Hz to thousands of Hz.
Despite the recent efforts in high frequency range~\cite{Fired_dataset, NILM_compressive}, and the intuition that sampling the load profiles at high rates 
provides more useful features for NILM, collecting this data requires special hardware.
The smart meter devices that are being installed widely in distribution grids 
can only provide low resolution data.
A recent study concludes that the data sampled at 1/30~Hz would be sufficient 
to achieve a high accuracy in NILM~\cite{NILM_resolution}. 
It is also mentioned that such a low sampling rate 
allows the NILM algorithms to benefit from additional information pertaining 
to the past power consumption of some long-running appliances.
A NILM method is proposed in~\cite{lowF2019} that is suitable for low-resolution data.
The performance of this method is evaluated for a range of resolutions, from 5~minutes to 1~hour.

Since combining the temporal information with active power measurements 
can facilitate disaggregation~\cite{7RayNILM}, 
most recent work utilizes Hidden Markov Model (HMM) and Deep Neural Networks (DNN).
Reference~\cite{15RayNILM} studies four variants of HMM for energy disaggregation using low frequency data.
Another low frequency power disaggregation method is proposed in~\cite{19RayNILM} 
that is based on HMM and DNN.
Reference~\cite{NeuralNILM} studies different deep learning architectures for NILM.
These methods can only generalize to unseen homes if the training set has enough variety,
which is rare in NILM training sets.
Reference~\cite{Seq2Point} proposes a sequence-to-point learning via a Convolutional Neural Network~(CNN)
that outperforms the sequence-to-sequence method introduced in~\cite{NeuralNILM}.
Reference~\cite{edgeNILM2020} studies different DNN compression schemas and 
suggests a multi-task learning-based architecture to further compress the models.

The majority of NILM methods assume that the net load is simply the total power consumed by home appliances.
There is just a few NILM papers that consider BTM solar generation.
Reference~\cite{NILMsolar2020} proposes a method to disaggregate three types of loads, 
namely traditional loads, distributed generation (e.g., PVs), and flexible loads (e.g., electric vehicles),
at the substation level.
However, the focus of our work is on customer-level solar disaggregation.
The closest line of work to ours is by Dinesh et. al~\cite{Dinesh2017}, 
which proposes a NILM method for customers with BTM solar generation.
The authors construct a unique set of signatures for appliances and solar generation,
and classify their operating modes using a spectral clustering based method.
Finally, they identify their state and operating mode through a subspace component power level matching algorithm.
The main drawback of this approach is that it relies on the synthesized net load measured at 1~second intervals.
This is much faster than the sampling rate of smart meters that are currently installed
in many jurisdictions (which is typically in the order of minutes).
\section{Problem Definition}
Customer-level solar disaggregation concerns decomposing the customer's net load, 
presumably measured by a smart meter, 
into home load and solar generation.
Since historical (disaggregated) data regarding solar generation and home load 
may not be available from a customer with a BTM PV system,
the problem should be solved in an unsupervised or semi-supervised fashion.

\subsection{Notation}
As a general rule, matrices and vectors are denoted respectively 
by bold face uppercase and lowercase letters.
We use a subscript to refer to a specific row of a matrix or an element of a vector.

Let $\textbf{y} \in \mathbb{R}^{T}$ be a vector that collects
the measured net load of one customer over $T$ intervals.
Hence, $\textbf{y}_{t}$ denotes the net load measurement of this customer at time $t \in \{1,2,...T\}$.
Similarly, let $\hat{\bm{\ell}}, \hat{\textbf{s}} \in \mathbb{R}^{T}$ denote respectively 
their estimated home load and behind-the-meter solar generation in the same period.
These quantities must satisfy the following equality constraint:
$\textbf{y} = \hat{\bm{\ell}} - \hat{\textbf{s}}$.

Let $\textbf{X}^L \in \mathbb{R}^{T\times K_l}$ be the set of $K_l$ features that determine the customer's home load,
and $\textbf{X}^S \in \mathbb{R}^{T\times K_s}$ be the set of $K_s$ proxy measurements 
that can be used to approximate the customer's solar generation via a mixture model.
We can train a non-linear model $g$ to map the features to the home load, 
and a linear solar mixture model to estimate the customer's solar generation. 
The model $g$ can be a neural network, a support vector machine, a random forest, 
or any other nonlinear model used for regression.
We can write:
\begin{align}
\hat{\bm{\ell}} &= g(\textbf{X}^L; \bm{\theta}), \\
\hat{\textbf{s}} &= \textbf{X}^S\textbf{w}, \label{eq:mixture}
\end{align}
where $\bm{\theta}$ is a vector that represents parameters of the load model, and
$\textbf{w}\in\mathbb{R}^{K_s}$ is the weight vector of the mixture model.
Hence, $\textbf{w}_{k}$ is a scalar that represents the weight assigned to the $k^\text{th}$ proxy.
The motivation for using a linear mixture model for estimating solar generation 
is explained in Section~\ref{subsection:solar_disaggregation}.


\subsection{Assumptions}
We postulate that no information is available about the customers 
except their approximate location (i.e., the city or district they are located in) and their smart meter data.
Hence, we do not know the exact longitude and latitude information of each customer.
This is a reasonable assumption because (a) some customers are reluctant to provide their home address,
and (b) there might be data privacy requirements that prevent the utility from sharing their address with
a third party that performs solar disaggregation or NILM.
In addition to the smart meter data, 
solar irradiance, wind speed, and ambient temperature 
at the city scale can be downloaded via an API.
We assume that deployment characteristics of BTM PV systems are not known a priori.
These include the panel size, orientation, tilt, temperature coefficient, etc.

We assume that there is at least one separately metered solar installation in the same city or district as our target home. 
This site provides proxy measurements that are used to estimate solar generation of the target home.
The deployment characteristics of this site could differ from the characteristics of the PV system installed at the target home.
We argue that this is not a strong assumption as utilities usually have access 
to direct solar measurements from several sites in a city.
Moreover, the cost of installing and maintaining one or a few PV systems can be easily justified 
by the potential benefits of observing customer-level PV generation.
\section{methodology}
We now introduce the models we use for estimating solar generation and home load.
We present our solar disaggregation algorithm, which has two main parts: 
a weight initialization technique and an iterative algorithm for updating the model parameters.

\subsection{Models}
\label{subsection:solar_disaggregation}
\begin{figure*}[ht]
  \includegraphics[width=\textwidth, height=3cm]{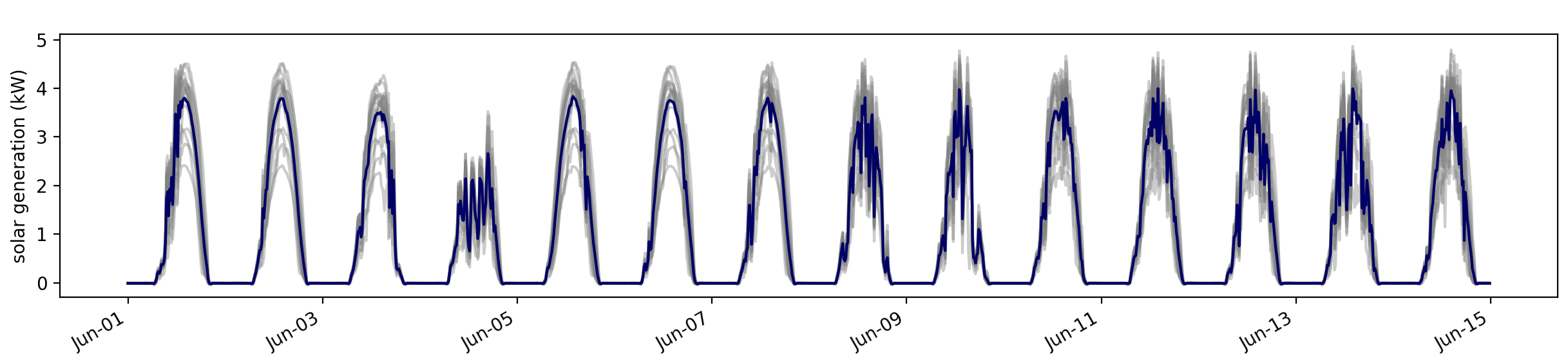}
  \vspace{-15px}
  \caption{The solar generation of 19 homes in the Pecan Street dataset over 2 weeks. 
  The thicker line shows the average solar generation across all the homes.}
  \label{fig:real_solar}
\end{figure*}

\textbf{Solar mixture model:} 
We aim to approximate that solar power generated by the BTM PV system 
installed at the target home using a mixture of proxy measurements from PV systems 
located in the same city or district. 
The intuition behind this approximation is that PV systems in the same geographical area 
have more or less the same solar generation pattern regardless of their deployment characteristics.
This can be verified by inspection of Figure~\ref{fig:real_solar} 
which displays the solar power generated by 19 homes with BTM PV systems in Austin, Texas.
The PV systems that have the same orientation but different sizes and tilts
have almost identical solar generation patterns, though with different scales.

There are two specific challenges that must be addressed to get a good approximation.
First, we do not have control over the deployment characteristics of 
the PV systems that provide proxy measurements.
If they had exactly the same orientation angle as the PV system installed at the target home,
estimating the target home's solar generation would reduce to learning a single scaling factor.
One way to address this challenge is to adopt a solar mixture model 
to approximate the target home's solar generation 
as a weighted sum of a number of proxy measurements, as shown in Equation~(\ref{eq:mixture}).
This increases the chance of getting proxy measurements from PV systems 
with similar deployment characteristics to the target home.
A higher weight will be eventually assigned to these PV systems in the mixture model.
The second challenge is that proxy measurements 
from a large number of neighbouring PV systems might not be available in practice.
To address this challenge, we combine proxy measurements from real PV systems 
with measurements synthesized by a physical PV model
that takes into account solar irradiance data of an arbitrary location in the same city.
We discuss in Section~\ref{sec:results} that it is essential to use proxy measurements 
from at least one real PV system as some fluctuations in solar generation
cannot be accurately explained by the physical PV model 
when coarse-grained solar irradiance data is fed to this model.
We also demonstrate the importance of incorporating synthetic proxies 
that have different orientation angles in Figure~\ref{fig:weight_assign}.
These synthetic proxies help us estimate the peak time of solar generation more accurately.

The physical PV model that we use to obtain data for \emph{synthetic proxies} is based on 
PVWatts~\cite{PVWatts}. We develop this model using the PV Performance Modeling Collaborative~\cite{PVPMC}.
The output power of the PV model with the specified rating $P_{dc0}$
can be computed given the transmitted plane of array (POA) irradiance $I_{tr}$ and cell temperature $T_{cell}$:
\begin{align}
P_{dc} = \frac{I_{tr}}{E_{ref}}P_{dc0}(1+\gamma(T_{cell} - T_{ref}))
\label{fun:PVWatts_model}
\end{align}
Here $\gamma$ represents the temperature coefficient, 
$E_{ref}$ represents the reference irradiance,
and $T_{ref}$ represents the reference cell temperature. 
We set them respectively to -0.47\%/$^{\circ}C$, 1000W/$m^2$, and 25$^{\circ}C$ to create synthetic proxies.
$I_{tr}$ is determined by solar irradiance data (direct normal irradiance, diffuse horizontal irradiance, global horizontal irradiance), PV system characteristics (e.g., tilt, orientation) and its location.
$T_{cell}$ is a function of wind speed, ambient air temperature, and solar irradiance data. 
We use different preset technical parameters for different synthetic proxies.

\textbf{Home load model:} To estimate the home load, we adopt a random forest regression, 
which is a supervised learning algorithm that uses an ensemble learning method for regression.
We use the scikit-learn~\cite{scikit} library to train this model.
Four explanatory variables are used as features, $\textbf{X}^L$, for all customers. 
These variables include ambient temperature $\textbf{c}$, 
exponentially weighted moving average of temperature over the last 24 hours $\textbf{c}_{wmv}$, hour of the day $\textbf{h}$, 
and a binary variable $\textbf{d}$ that indicates if it is a weekday or weekend. 
Thus, we have $\textbf{X}^{L} = [\textbf{c}, \textbf{c}_{wmv}, \textbf{h}, \textbf{d}]$.

\subsection{Solar Disaggregation}
 
\textbf{Weight initialization:} 
The first step for implementing our method is to initialize the weight vector $\textbf{w}$ of the solar mixture model
using the net load data of the target home and the solar generation data collected from the proxies.
A good initialization can enhance the performance of the disaggregation method and reduce its convergence time.
Our weight initialization method has 3 main steps:
\begin{enumerate}
  \item Estimating the physical characteristics of PV systems installed at the target home and real solar proxies.
  \item Finding the maximum solar generation of each PV system.
  \item Solving an optimization problem to determine the initial weight vector $\textbf{w}$.
\end{enumerate}
We use an open source toolkit, SolarTK~\cite{bashir2019},
to estimate the physical characteristics of PV systems including its tilt, orientation, and panel size.
To estimate these parameters, the toolkit takes the real solar generation data 
as input and finds the maximum solar generation.
We can run this toolkit on proxy measurements, but we lack the real solar generation data from the target home.
To solve this problem, we approximate the solar generation of the target home given the net load data $\textbf{y}$ 
from this expression $\hat{\textbf{s}} \approx [\ell_{base} - \textbf{y}]^+$, 
where $[~~]^+$ is an operator that truncates negative elements of a vector to zero, 
and $\ell_{base}$ is the target home's base power consumption calculated as the minimum consumption level at night time.
SolarTK is then applied to the estimated solar generation of the target home and 
the real solar generation of solar proxy/proxies 
to obtain the estimated parameters for all PV systems.
Since we use a city's longitude and latitude as an approximate location for all the PV systems located in it, 
the estimated parameters may not be highly accurate.

We then calculate the maximum solar generation for each proxy and target home 
using the estimated deployment characteristics obtained in Step~1.
The maximum solar generation is the potential generation of a PV system under clear sky condition, 
that is determined by the system's physical characteristics, the ambient temperature and the location of PV.
We denote the maximum solar generation of the $k^{\text{th}}$ proxy by $\textbf{m}^{p}_{k}\in\mathbb{R}^{T}$,
and the maximum solar generation of target home by $\textbf{m}^{c}\in\mathbb{R}^{T}$.

In the last step, we determine the initial weight vector, $\textbf{w}$, for the solar mixture model
following the idea of~\cite{Shaker2016,Sossan2018};
the solar generation of a target home with unknown deployment characteristics is estimated
utilizing metered solar generation of sites with nonuniform deployment characteristics.
Formally, we can write
\begin{align} 
\textbf{s}^{target}_t = \alpha \cdotp \textbf{s}^{proxy}_t
\label{fun:transposition_in_our_format}
\end{align}
where $\alpha$ depends on time, site location, and other site-specific factors.
In our method, we simplify $\alpha$ to be a constant weight factor for each site, i.e., $\textbf{w}_k$.
Since we do not have the true solar generation from target home in Equation~(\ref{fun:transposition_in_our_format}),
we use the maximum solar generations to determine the initial weight of each solar proxy.
Specifically, the weight factor $\textbf{w}_k$ for the $k^{\text{th}}$ proxy 
can be determined by solving the following optimization problem: 
\begin{equation}
\begin{aligned}
&\min_{\textbf{w}_k} & \| \textbf{w}_k\cdotp\textbf{m}^p_k - \textbf{m}^{c}\|_2 \\
&\textrm{subject to} & \textbf{w}_k > 0  
\label{fun:determine_weight}
\end{aligned}
\end{equation}

\begin{algorithm}[t!]
\caption{Solar disaggregation for one target customer}
\label{alg:solar_dis}
\SetAlgoLined
\SetKwInOut{Input}{Input}\SetKwInOut{Output}{Output}
\SetKwInput{Init}{Init}
\Input{
Net load of the target customer, $\textbf{y} \in \mathbb{R}^{T}$\; 
\ \ \ \ \ \ \ \ \ \ \ \ \ \ \ Proxy measurements from $K$ sites, $\textbf{X}^S \in \mathbb{R}^{T\times K}$\;
\ \ \ \ \ \ \ \ \ \ \ \ \ \ \ Initial weights of the solar mixture model, $\textbf{w} \in \mathbb{R}^{K}$\;
\ \ \ \ \ \ \ \ \ \ \ \ \ \ \ Load related features, $\textbf{X}^L$\;
}
\Output{Estimated solar generation and home load of the target customer, $\mathbf{\hat{\textbf{s}}}$, $\mathbf{\hat{\bm{\ell}}}$;}
\Init{
$\textbf{w}^{0} \leftarrow \textbf{w} / K$\;
\ \ \ \ \ \ \ \ \ \ $\textbf{s}^{0} \leftarrow \textbf{X}^S\textbf{w}^{0}$\; 
\ \ \ \ \ \ \ \ \ \ Initialize parameters $\bm{\theta}^{0}$ for load model $g$\;
}
 \While{$iter < Max \ Iteration$ \textbf{and} $|\textbf{w}^{iter} - \textbf{w}^{iter-1}| > \epsilon$}{
    $\bm{\ell}^{iter} \leftarrow \textbf{s}^{iter} + \textbf{y}$\;
    Incrementally train the model g with input feature $\textbf{X}^{L}$ and output $\bm{\ell}^{iter}$\;
    Update load $\bm{\ell}^{iter} \leftarrow g(\textbf{X}^{L},\bm{\theta}^{iter})$\;
    $\textbf{s}^{iter} = \bm{\ell}^{iter} - \textbf{y}$\;
    $\textbf{w}^{iter} \leftarrow  \argmin_{\textbf{w}} \|\textbf{X}^S\textbf{w} - \textbf{s}^{iter}\|_2$\;
    $\textbf{s}^{iter} \leftarrow \textbf{X}^S\textbf{w}^{iter}$\; 
    }
\end{algorithm}
\textbf{Disaggregation algorithm:} 
In this step, 
we iteratively estimate the home load and solar generation until the parameters of our model converge.
Algorithm~\ref{alg:solar_dis} presents the pseudocode of the proposed solar disaggregation algorithm.
After obtaining the initial weights $\textbf{w}$ for the solar mixture model, 
we first estimate the solar PV generation $\textbf{s}^{iter}$ using a linear combination of the solar proxies.
Then, we use $\textbf{y} = \hat{\bm{\ell}} - \hat{\textbf{s}}$ to calculate the estimated home load $\bm{\ell}^{iter}$ (line 2) and incrementally train the load model using $\bm{\ell}^{iter}$ and load related features $\textbf{X}^L$ (line 3).
Based on the updated home load $\bm{\ell}^{iter}$ (line 4), we determine solar generation $\textbf{s}^{iter}$ (line 5), 
update the weights for solar proxies (line 6), and recalculate the solar generation using the updated weights (line 7). 
We repeat the above steps until the solar proxy weights $\textbf{w}$ converge or we reach the maximum number of iterations.
In our experiments, it typically takes between 20 and 80 iterations for this algorithm to converge
depending on the number of proxies and goodness of initial weights.
\section{Evaluation}

\begin{figure}[t!]
\centering
\includegraphics[width=0.65\linewidth]{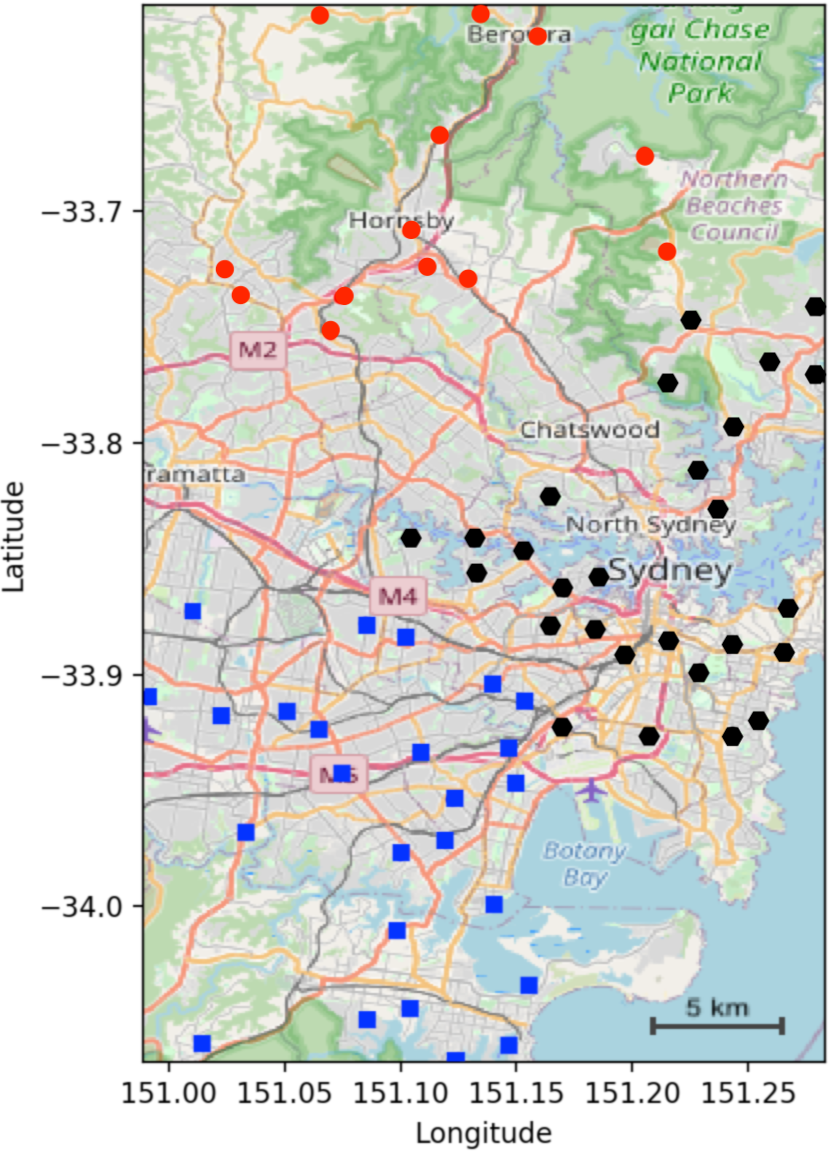}
\vspace{-8px}
\caption{The approximate locations of customers in Ausgrid dataset. The unit of scale bar is 5 km.}
\label{fig:map}
\end{figure}

\subsection{Datasets}
We use two publicly available datasets, 
namely Pecan Street~\cite{pecan2018} and Ausgrid~\cite{ausgrid},
to evaluate the estimation accuracy of different solar disaggregation methods.
Both datasets include net load measurements in addition to direct measurements of
home load and solar generation from homes with rooftop PV systems.

\textbf{Pecan Street dataset:} This dataset contains customer-level measurements 
of solar generation and appliance loads from homes that are located 
in different cities in New York, California, and Texas.
Since the target home and solar proxies need to be in the same city,
we restrict our focus to 19 homes that are located in Austin, Texas (in the northern hemisphere),
and have solar generation and home load data.
We set the latitude and longitude of all PV systems to 30.292432 and -97.699662, 
which are the coordinates of a randomly selected location in Austin.
We collect 15-minute interval data from each home over two time periods, 
one in the summer and one in the winter.
Specifically, the summer period ($T$=2880) is from June 1, 2018 to June 30, 2018 
and the winter period ($T$=2688) is from December 3, 2018 to December 30, 2018.

\textbf{Ausgrid dataset:} This dataset consists of 150 customers with rooftop PV systems in Sydney, Australia 
(in the southern hemisphere) with the latitude and longitude of -33.888575 and 151.187349 respectively. 
Figure~\ref{fig:map} shows the locations of these customers\footnote{The map is downloaded from OpenStreetMap~\cite{OpenStreetMap}.}.
These locations are approximate since we only have the postal code of each customer. 
It is also likely that a marker in this figure represents multiple customers that have the same postal code.
Since Sydney is a sprawling city, 
we cluster the customers into three clusters according to their latitude and longitude.
Points within each cluster are drawn in the same color and shape in Figure~\ref{fig:map}. 
As it can be seen, each cluster still spans a large area of the city.
Unlike the data obtained from Pecan Street,  
the solar generation and home load in this dataset are sampled at 30-minute intervals.
We consider two periods in two seasons, 
one from November 1, 2012 to November 30, 2012 in the summer season ($T$=1440) 
and the other one from May 1, 2013 to May 30, 2013 in the winter season ($T$=1440). 
A small number of customers are removed due to data quality issues in each season.

\textbf{Weather data:}
For the city of Austin, we acquire solar radiation, wind speed, and outside air temperature
from the National Solar Radiation Database~\cite{NSRD}.
This gives us one reading every 30 minutes.
To match the temporal resolution of the load data,
we use linear interpolation to upsample the weather data (to one reading every 15 minutes).
For the city of Sydney in Australia, we pull in the respective weather and irradiance data 
with 30-minute temporal resolution using the Solcast API~\cite{solcast}.
The load data has the same temporal resolution in this case.

\subsection{Variants of our Disaggregation Method}
\label{subsection:method_variations}
\begin{table}[t]
\centering
\caption{Physical parameters of synthetic proxies in different proxy settings.}
\begin{tabular}{lp{0.03\textwidth}lp{0.03\textwidth}lp{0.03\textwidth}l}
\toprule
& \multicolumn{2}{c}{orientation ($^{\circ}$)} & \multicolumn{2}{c}{tilt ($^{\circ}$)} & \multicolumn{2}{c}{DC rating (kW)}\\
\cmidrule(r){2-3}\cmidrule(r){4-5}\cmidrule(r){6-7}
Method & {Pecan} & {Ausgrid}  & {Pecan} & {Ausgrid} & {Pecan} & {Ausgrid} \\
\midrule
1P+1SP(SP1) & 180 & 0 & 30.3 & 33.9 & 3 & 3\\
1P+3SP(SP1) & 180 & 0 & 30.3 & 33.9 & 3 & 3\\
1P+3SP(SP2) & 90 & 90 & 30.3 & 33.9 & 3 & 3\\
1P+3SP(SP3) & 270 & 270 & 30.3 & 33.9 & 3 & 3\\
\bottomrule
\end{tabular}
\footnotesize{1P+3SP(SPx) refers to the $x^\text{th}$ synthetic proxy in 1P+3SP; 
The panel faces respectively N, E, S, and W when the orientation angle is $0^{\circ}$, $90^{\circ}$, $180^{\circ}$, $270^{\circ}$.}\
\label{tab:proxy_setting}
\end{table}

We implement our method with 4 different solar proxy settings.
\begin{itemize}
  \item \textbf{3Proxies: } we directly use solar generation data for the same periods 
  from 3 real rooftop PV systems in the same city.
  \item \textbf{1P+1SP: } we only use 1 real solar proxy combined with 1 synthetic proxy. 
  In this case, the ideal orientation angle in each hemisphere is used to create the synthetic proxy.
  It is specifically $180^{\circ}$ in the northern hemisphere and $0^{\circ}$ in the southern hemisphere.
  \item \textbf{1P+3SP: } we use 1 real solar proxy combined with 3 synthetic proxies 
  with different orientation angles.
  \item \textbf{3SP: } we use 3 synthetic proxies with different orientation angles just like the previous setting.
\end{itemize}
The parameters for different synthetic proxies are shown in Table~\ref{tab:proxy_setting}.
We set the tilt angle to the absolute value of the city's latitude 
and use a uniform DC rating for all synthetic proxies.
The tilt and DC rating have a similar effect on solar generation curve, 
i.e., they scale the curve up or down~\cite{Chen2017},
whereas the orientation shifts the peak of the generation curve to earlier or later.
Therefore, we can set the tilt angle and DC rating similarly for all the synthetic solar proxies 
because the elements of our weight vector $\textbf{w}$ will be adjusted by Algorithm~\ref{alg:solar_dis}.

\subsection{Baselines} 
We compare the performance of our solar disaggregation method with 
two methods that also use the data that is commonly available to the utility 
and outperform other solar disaggregation methods proposed in the literature.
Specifically, we use the solar disaggregation methods proposed in \cite{Kabir2019} and \cite{bashir2019} as our baselines;
these methods are labelled ``Baseline 1'' and ``Baseline 2'', respectively.
For a fair comparison, we implement the one-nearby-proxy-based solar estimation method in \cite{bashir2019} 
that was adopted in the case study of computing the clear sky index.
In each experiment, we use the same real solar proxy for our method and Baseline~2.
It is worth mentioning that in some homes the PV system's physical characteristics 
estimated by Baseline~2 can be quite different from their true values
due to the inaccuracy of location information and lack of directly measured solar generation from the respective homes.
Therefore, in our implementation of Baseline~2 we constrain the system's orientation 
within specific bounds based on the hemisphere in which the home is located.

\subsection{Evaluation Metrics}
We use two metrics to assess the performance of our disaggregation method with different proxies,
and compare them with the two baselines.
The first metric is the root-mean-square error (RMSE) which has been used 
in many solar disaggregation papers.
The second one is a normalized RMSE metric, called nRMSE.
Equation~(\ref{fun:CV}) is the definition of nRMSE for the estimated solar generation of a customer.
It is the RMSE normalized by the mean value of the real solar generation.
\begin{equation}
\begin{split}
nRMSE &= \frac{\sqrt{\sum_{t=1}^{T}(\textbf{s}_t - \hat{\textbf{s}}_t)^2 / T}}{\sum_{t=1}^{T}\textbf{s}_{t} / T} = \frac{\text{RMSE}}{\text{Mean}}\\
\end{split}
\label{fun:CV}
\end{equation}
Compared to RMSE, this normalized metric can help us compare the disaggregation performance 
on signals with different magnitudes (i.e., generation from PV panels with different sizes).

\renewcommand\arraystretch{1.3} 
\begin{table*}[t]
\centering
\caption{Comparison of various disaggregation methods. Each cell contains two slash-separated metrics: nRMSE and RMSE.}
\begin{tabular}{lp{0.09\textwidth}p{0.09\textwidth}p{0.09\textwidth}p{0.09\textwidth}p{0.09\textwidth}p{0.09\textwidth}p{0.09\textwidth}p{0.09\textwidth}}
\toprule
& \multicolumn{4}{c}{Summer} & \multicolumn{4}{c}{Winter}\\
\cmidrule(r){2-5}\cmidrule(r){6-9}
& \multicolumn{2}{c}{Ausgrid} & \multicolumn{2}{c}{Pecan} & \multicolumn{2}{c}{Ausgrid} & \multicolumn{2}{c}{Pecan}\\
\cmidrule(r){2-3}\cmidrule(r){4-5}\cmidrule(r){6-7}\cmidrule(r){8-9}
Method & {Solar} & {Load} & {Solar} & {Load} & {Solar} & {Load} & {Solar} & {Load}   \\
\midrule \midrule
3Proxies & 0.469/0.0621 & 0.232/0.0593 & 0.357/0.359 & 0.157/0.353 & 0.771/0.0602 & 0.199/0.0582 & 0.300/0.142 & 0.164/0.136\\
1P+1SP & 0.543/0.0723 & 0.273/0.0686 & 0.484/0.489 & 0.211/0.474 & 0.841/0.0677 & 0.226/0.0649 & 0.375/0.179 & 0.211/0.175\\
1P+3SP & 0.525/0.0691 & 0.254/0.0650 & 0.510/0.515 & 0.223/0.500 & 0.797/0.0638 & 0.212/0.0617 & 0.336/0.161 & 0.189/0.158\\

3SP & 0.643/0.0858 & 0.305/0.0780 & 0.604/0.610 & 0.259/0.581 & 0.854/0.0702 & 0.231/0.0669 & 0.420/0.204 & 0.239/0.198\\

Baseline 1 & 0.612/0.0778 & 0.287/0.0778 & 0.516/0.493 & 0.221/0.493 & 1.713/0.1210 & 0.375/0.1210 & 0.516/0.254 & 0.284/0.254\\
Baseline 2 & 0.631/0.0857 & 0.337/0.0857 & 0.518/0.478 & 0.224/0.478 & 0.927/0.0767 & 0.272/0.0767 & 0.508/0.232 & 0.280/0.232\\
\bottomrule
\end{tabular}
\label{tab:result_comparison}
\end{table*}

\begin{figure}[h]
\begin{subfigure}{\linewidth}
  \centering
  \includegraphics[width=\textwidth]{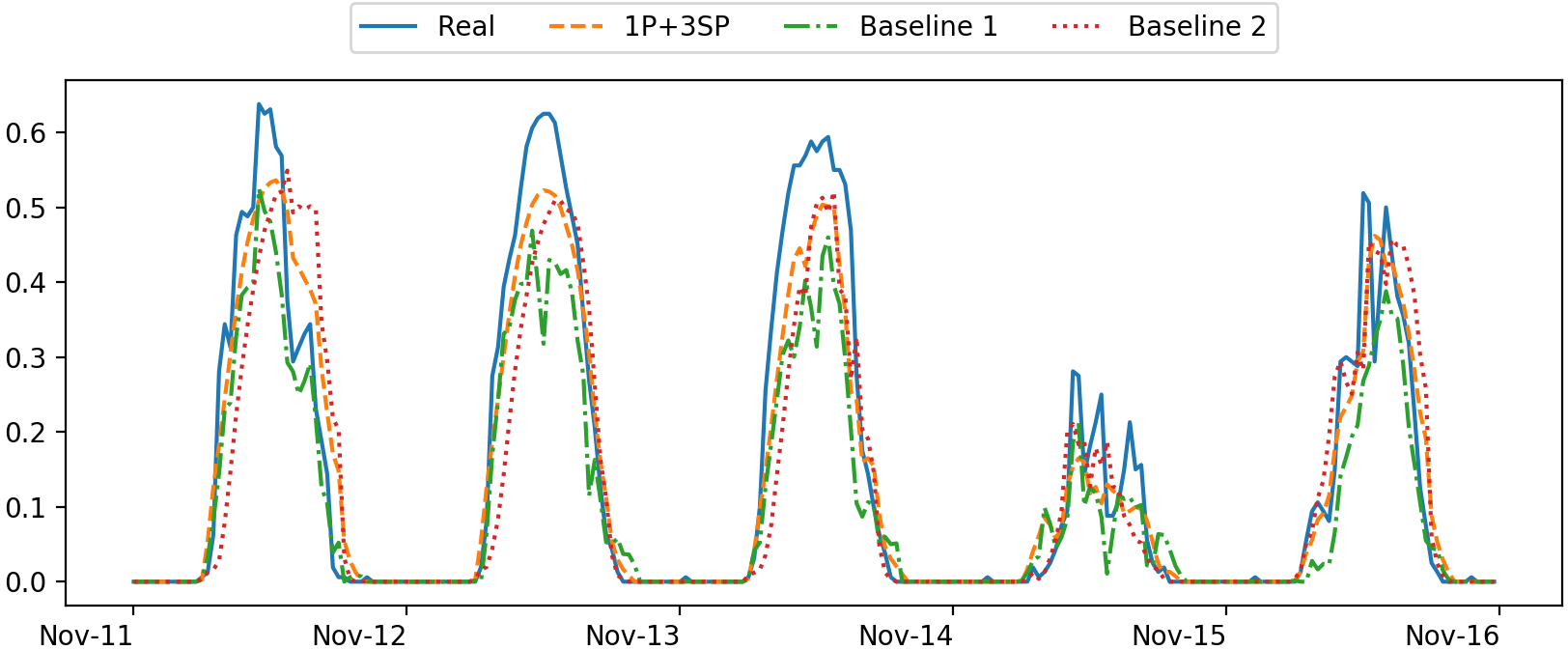}
\end{subfigure}
\begin{subfigure}{\linewidth}
  \centering
  \includegraphics[width=\textwidth]{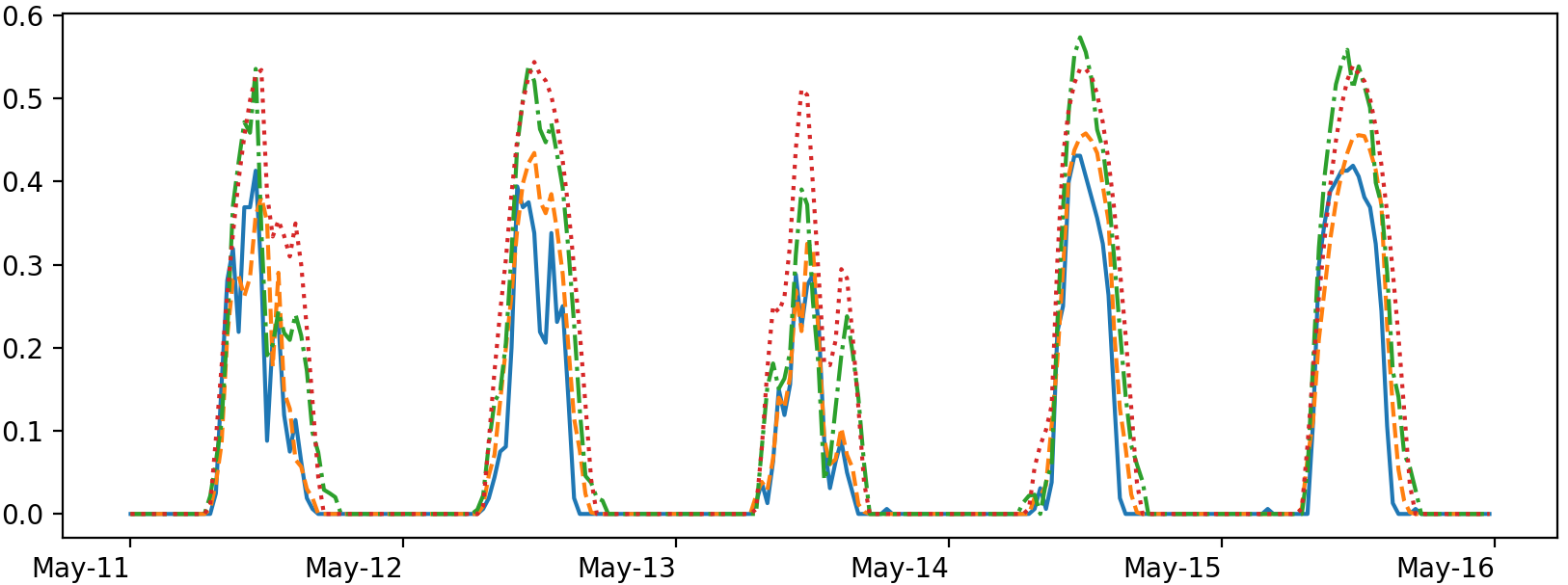}
  \vspace{-15px}
\end{subfigure}
\caption{Comparison of disaggregated solar generation in summer (top) and winter (bottom) for a customer. 
}
\label{fig:home_disaggregation_plot}
\end{figure}

\section{Experimental Results}\label{sec:results}
We first evaluate the performance of our methods in disaggregating BTM solar generation
using the metrics introduced in the previous section.
We analyze the sensitivity of our methods to the amount of net meter data used for disaggregation,
the choice of solar proxies, and the weight initialization method.
Lastly, we investigate the impact of running NILM methods 
on the disaggregated solar generation, real solar generation, and net load.
This will reveal the potential benefits of disaggregating solar generation prior to performing NILM.

\begin{figure*}
        \centering
        \begin{subfigure}[b]{.49\linewidth}
            \centering
            \includegraphics[width=\textwidth,
            height=4cm]{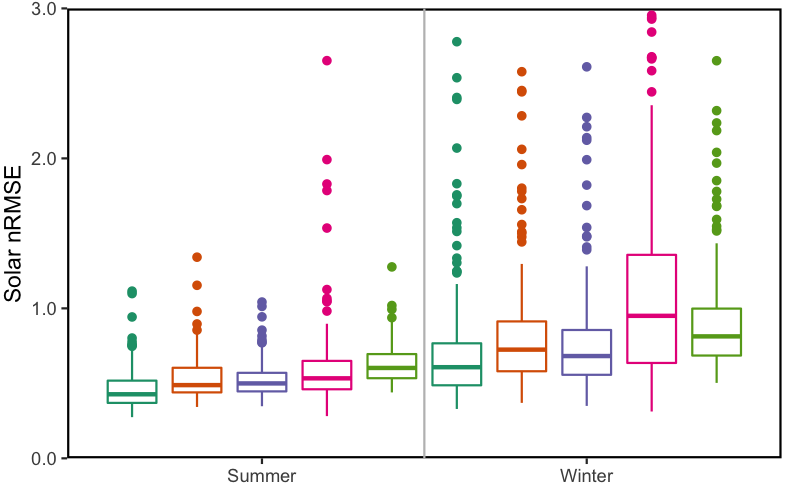}
        \end{subfigure}
        \hfill
        \begin{subfigure}[b]{.49\linewidth}  
            \centering 
            \includegraphics[width=\textwidth, height=4cm]{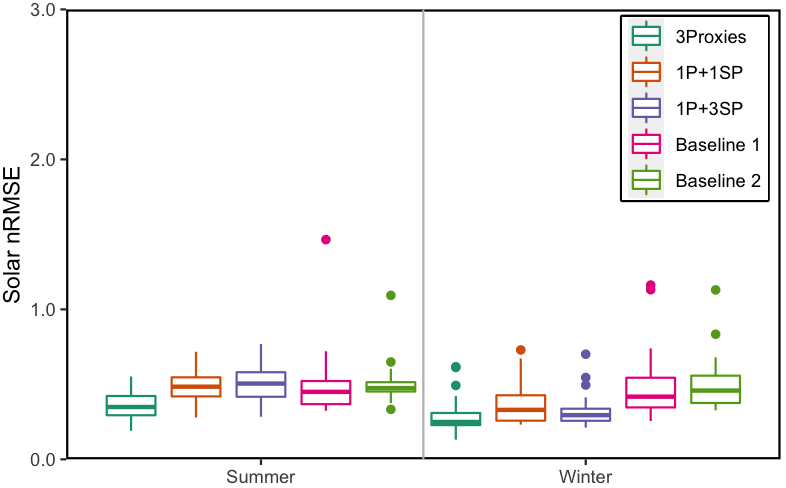}
        \end{subfigure}
        \caption{The distribution of nRMSE values for the estimated solar generation  
        in Ausgrid (left column) and Pecan Street (right column) datasets. 
        The legend shows the solar disaggregation methods shown in each panel (from left to right). The length of the whiskers is 1.5 times IQR.} 
        \label{fig:result_comparison}
    \end{figure*}


\begin{figure}[ht]
\begin{subfigure}{\linewidth}
  \centering
  \hspace*{0cm}\includegraphics[width=\linewidth, height=3.5cm]{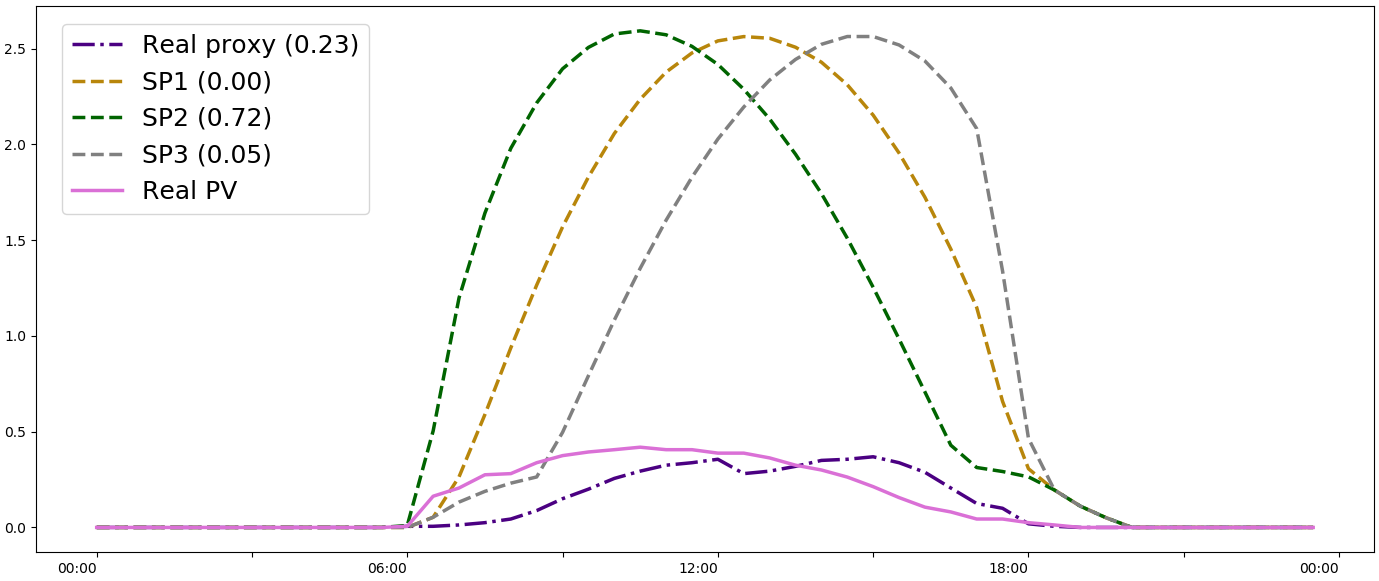}  
\end{subfigure}
\caption{PV generation (kW) of a sample home. Dashed curves show proxy measurements.
The relative weights (normalized to sum to 1) of proxies are put in the legend.
}
\label{fig:weight_assign}
\end{figure}

\subsection{Solar Disaggregation Performance}
We compare 4 variants of our method -- 3Proxies, 1P+1SP, 1P+3SP, and 3SP --
with the two baselines described in Section~\ref{subsection:method_variations}.
For each variant, we evaluate the disaggregation performance for all customers with PV systems in the dataset.
Since our method utilizes proxy measurements, 
we run the experiment 10 times for each target home 
with real solar proxies that are randomly selected from the pool (excluding the target home).
For the Ausgrid dataset, we select real solar proxies from the same cluster as the target home (shown in Figure~\ref{fig:map}).

\textbf{Estimation accuracy: } 
Table~\ref{tab:result_comparison} shows the average nRMSE and RMSE 
of solar generation and home load estimation
across all customers in two seasons and two different datasets.
We observe that 3Proxies yields the lowest error compared to the other variants of our method and the two baselines.
1P+1SP outperforms the two baselines in both seasons,
reducing nRMSE by 29.74\% and 12.14\% on average, respectively.
Similarly, 1P+3SP outperforms the two baselines in all cases except summer in the Pecan the Street dataset;
in this case, Baseline~1 and Baseline~2 have comparable performance to 1P+3SP.
Specifically, 1P+3SP has lower nRMSE for solar estimation,
while Baseline~1 and Baseline~2 have lower RMSE (on average) for solar and load estimation.
Overall, 1P+3SP reduces nRMSE of solar estimation by 32.31\% and 15.66\% on average 
compared to Baseline~1 and Baseline~2, respectively.
This observation suggests that by utilizing as few as only one directly measured solar generation site, 
we can get a better estimate of solar generation and home load than the state-of-the-art solar disaggregation methods.
Interestingly, 3SP has the worst performance among the four variants of our method,
although it still beats Baseline~1 and Baseline~2 in winter.
This underscores the importance of having at least one real proxy for solar disaggregation
to account for high frequency variations (e.g., due to a passing cloud) in the solar generation.
We do not consider 3SP in the next section for sensitivity analysis as it does not use a real proxy.

Figure~\ref{fig:home_disaggregation_plot} illustrates the disaggregated solar generation 
in two seasons by 1P+3SP and the two baseline methods for a randomly selected home from the Ausgrid dataset.
It can be seen that our estimated solar is generally closer to true solar generation in terms of estimating peak generation time and generation scale.
Figure~\ref{fig:result_comparison} shows box and whisker plots 
indicating the nRMSE distribution for each disaggregation method. 
All the methods have relatively small variances in the Pecan Street dataset,
though the variances are noticeably larger for the Ausgrid dataset, 
especially in winter where there are several outliers.
It can be seen that our method (1P+3SP) outperforms both baselines for most homes.
In a small number of homes, its performance is on par with Baseline~2 that also 
utilizes proxy solar measurements from a neighbouring site.
This suggests that using synthetic proxies in addition to the real proxy 
improves the estimation accuracy in most cases.

Figure~\ref{fig:weight_assign} shows real solar generations of a target home besides 
the four proxy measurements used in 1P+3SP.
It can be seen that the peak generation of the target home and real solar proxy happen at different times 
as they have different orientations.
In this case, the synthetic proxy with an orientation angle close to the target home's orientation angle 
gets a much higher weight compared to the other synthetic proxies and the real proxy.
This highlights the advantage of incorporating the synthetic proxies.

\textbf{Computation time: } 
We now compare our method (the 1P+3SP variant) with the baselines in terms of their running time.
Excessively high running times could be prohibitively costly
for the utilities that intend to run the solar disaggregation method at scale.
We use an Intel Xeon Silver 4114 CPU to run the experiments. 
The total running time for disaggregating one month data from a target home in Ausgrid using 1P+3SP 
is around 35~seconds (25~seconds for weight initialization and 10~seconds for solar disaggregation).
The other variants of our method have similar running times.
In comparison, 
the total running time of Baseline~1 
is about 70 minutes,
which is roughly 120 times greater than the running time of our method.
We attribute this to the fact that this method 
solves a nonlinear optimization problem for the solar model and 
trains a Markov switching regression model for load estimation in every iteration.
Although Baseline~2 has a running time that is similar to our proposed method (i.e., 30 seconds), 
its estimation accuracy is worse than ours as explained in the previous section.

\subsection{Sensitivity Analysis}
\begin{figure}[t!]
\centering
\includegraphics[width=\linewidth]{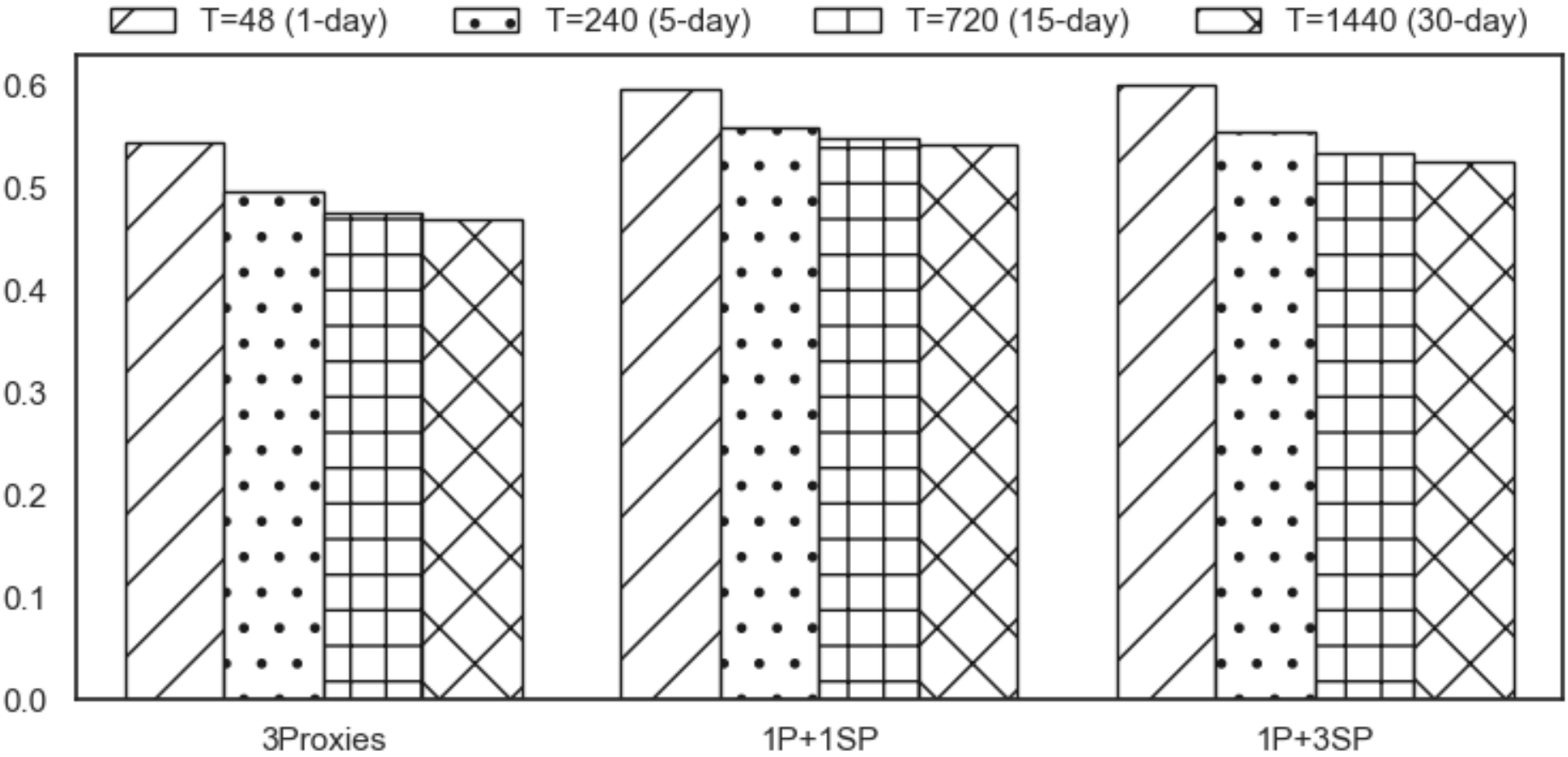} 
\caption{nRMSE of disaggregated solar for different disaggregation lengths in the Ausgrid dataset.}
\label{fig:sensitivity_length}
\end{figure}

\textbf{Disaggregation length:} 
We now investigate how extending the length of solar generation data would impact the performance of our disaggregation method.
We set the disaggregation length $T$ to be \{48, 240, 720, 1,440\} time intervals 
corresponding to \{1, 5, 15, 30\} days of data from Ausgrid.
Then, we apply our method to each $T$ consecutive intervals separately.
Similar to the implementation in the previous subsection, 
we run 10 independent experiments and average the error values for each target home 
to account for the randomness caused by choosing different solar proxies. 
Figure~\ref{fig:sensitivity_length} shows the nRMSE of 3Proxies, 1P+1SP, and 1P+3SP 
for different disaggregation length.
For all three variants, the nRMSE is larger for shorter disaggregation periods.
This could be because more data is available for longer periods,
which is helpful to capture the true relationship between the solar generation of the target home and solar proxies.

\begin{figure}[ht]
\begin{subfigure}{\linewidth}
  \centering
  \includegraphics[width=\linewidth, height=3cm]{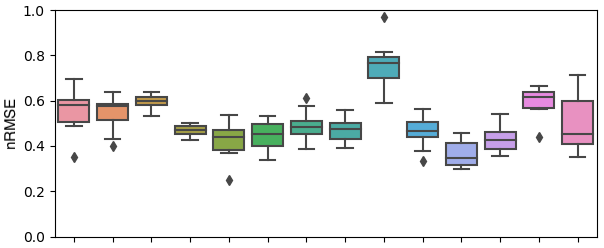}  
\end{subfigure}
\begin{subfigure}{\linewidth}
  \centering
  \includegraphics[width=\linewidth, height=3cm]{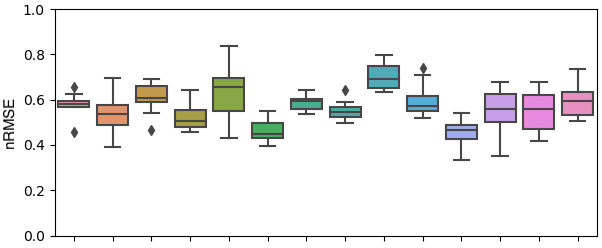}  
\end{subfigure}
\begin{subfigure}{\linewidth}
  \centering
  \includegraphics[width=\linewidth, height=3cm]{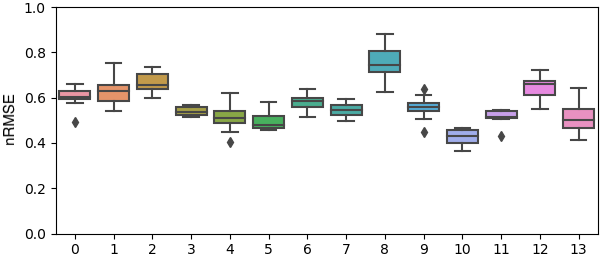}  
  \vspace{-15px}
\end{subfigure}
\caption{The distribution of nRMSE values obtained using different choices for the proxy measurement. The top plot shows the results from 3Proxies, the middle plot shows the results from 1P+1SP, and the bottom one shows the results from 1P+3SP. The x-axis indicates the home index.}
\label{fig:sensitivity_proxy_selection}
\end{figure}

\begin{figure}[ht]
\begin{subfigure}{\linewidth}
  \centering
  \hspace*{0cm}\includegraphics[width=\linewidth]{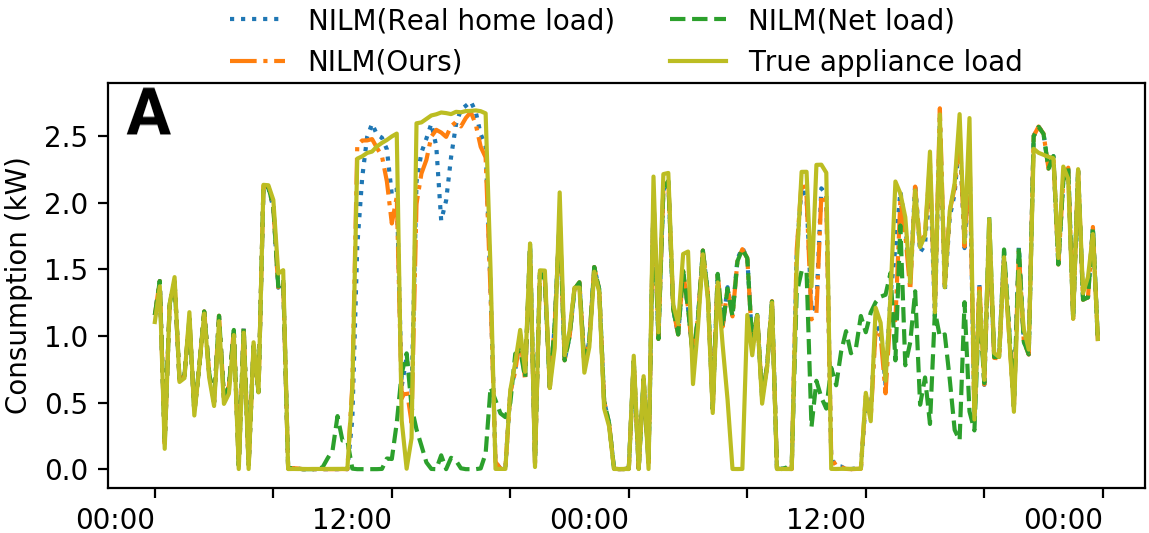}  
\end{subfigure}
\begin{subfigure}{\linewidth}
  \vspace{10px}
  \centering
  \hspace*{0cm}\includegraphics[width=\linewidth]{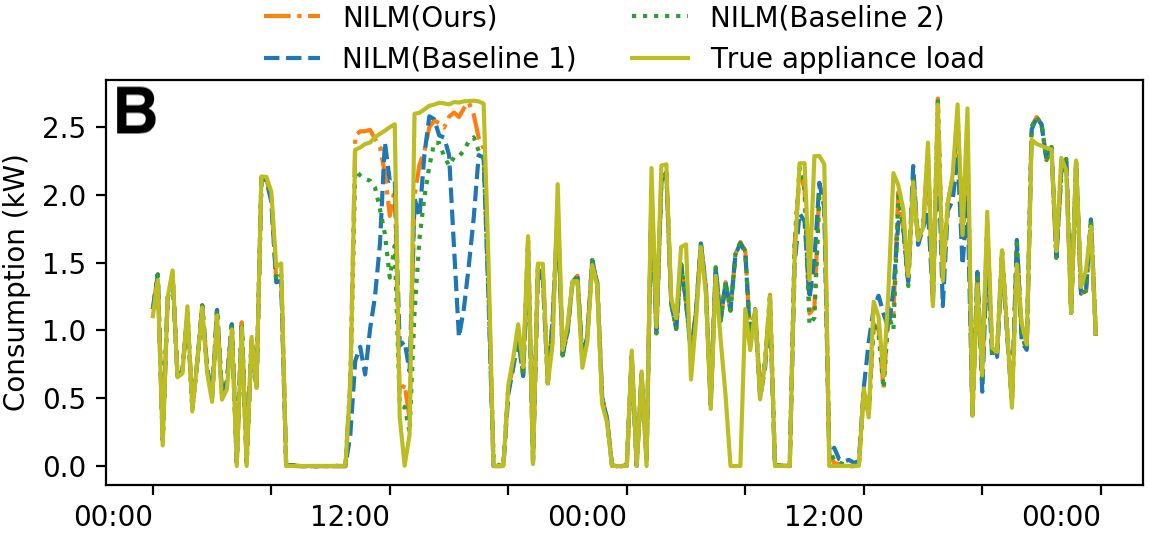}  
  \vspace{-10px}
\end{subfigure}
\caption{Comparison between the true and estimated air conditioner (AC) loads.
The top plot shows the results of running Seq2Point on real home load, net load, 
and the home load disaggregated by our method (1P+3SP). 
The bottom plot shows the results of running Seq2Point 
on the home load disaggregated by our method (1P+3SP) and 2 baselines.}
\label{fig:NILM_appliance}
\end{figure}

\textbf{Selection of solar proxies:} To evaluate the sensitivity of our methods to the choice of solar proxies,
we randomly choose 14 target homes in Ausgrid and run 10 independent experiments 
with solar proxies that are randomly chosen in each experiment.
Figure~\ref{fig:sensitivity_proxy_selection} shows the nRMSE distribution for the 14 different target homes.
It can be seen that the nRMSE distributions obtained for a few target homes are wider than the rest, 
implying that the disaggregation method is more sensitive to the choice of solar proxies.
This is because these homes are located far from the majority of homes in this dataset.
Expectedly, 1P+3SP is the least sensitive variant to the choice of solar proxy
because it only requires one real solar proxy and incorporates three synthetic proxies. 

\begin{table}[h]
\centering
\caption{Comparing nRMSE of disaggregated solar using different methods to initialise the weight vector $\textbf{w}$.}
    \begin{tabular}{p{2.2cm}p{1.5cm}p{1.5cm}p{1.5cm}}
    \toprule
    \rule{0pt}{10pt}
    & Constant & Random & Ours\\
    \midrule
    3Proxies & 0.78553 & 1.49346 & 0.46856\\
    1P+1SP & 2.80917 & 2.93639 & 0.54347\\
    1P+3SP & 3.80280 & 8.85139 & 0.52521\\
    \bottomrule
    \end{tabular}
\label{tab:initialization}
\end{table}

\begin{figure*}[ht]
\begin{subfigure}{\textwidth}
  \centering
  \includegraphics[width=.95\linewidth]{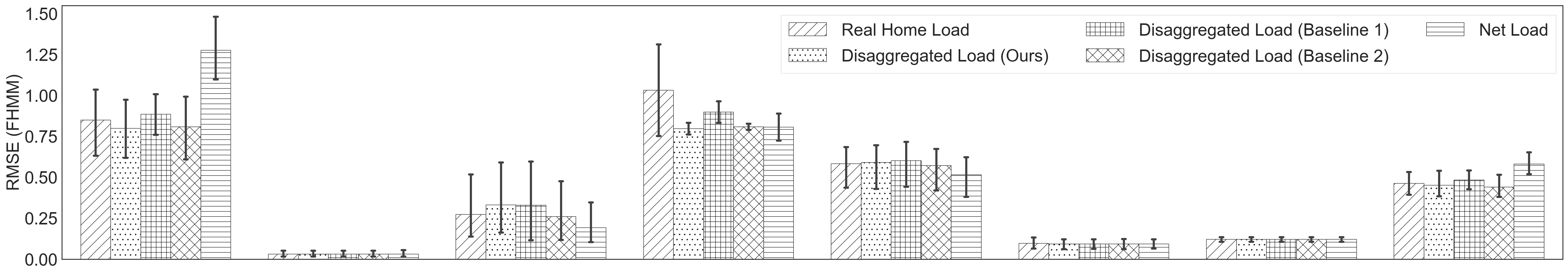}  
\end{subfigure}
\begin{subfigure}{\textwidth}
  \centering
  \includegraphics[width=.95\linewidth]{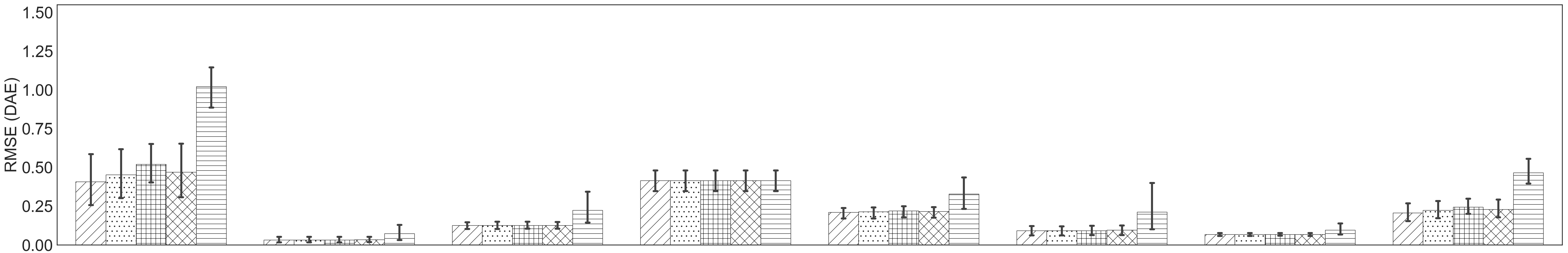} 
\end{subfigure}
\begin{subfigure}{\textwidth}
  \centering
  \includegraphics[width=.95\linewidth]{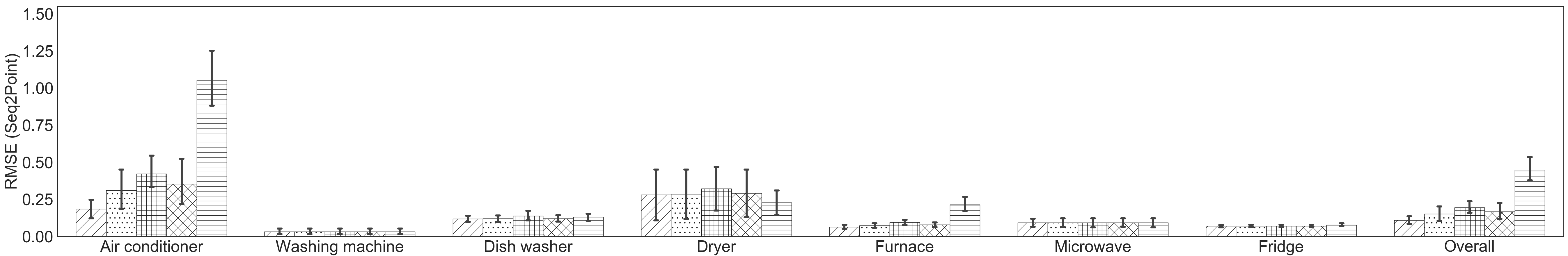}  
\end{subfigure}
\caption{Average RMSE of each appliance among all selected homes. The top plot shows the disaggregation performance of FHMM, the middle plot shows the performance of DAE, and the bottom one shows the performance of Seq2Point. 
Error bars show the 95\% confidence interval.}
\label{fig:NILM_average_error}
\end{figure*}

\textbf{Weight initialization:} 
To evaluate the efficacy of our initialization method, 
we compare it with constant and random initialization methods.
We simply set $\textbf{w}$ to $\mathbf{1}$ for constant initialization.
For random initialization, we assign random numbers in the range of $[0, 1]$ to the weights.
Table~\ref{tab:initialization} compares the results.
It is evident that our weight initialization method can significantly increase the accuracy of solar disaggregation, 
especially when we use only one real solar proxy (i.e., 1P+1SP and 1P+3SP).
We attribute this to the fact that we take into account the PV system deployment characteristics
estimated by SolarTK.

\subsection{NILM Performance}

We study the effect of BTM solar generation on the performance of NILM techniques.
For the purpose of this study, 
we select 6 homes from the Pecan Street dataset 
that have both solar generation and individual appliance consumption data,
and apply three benchmark NILM techniques, 
namely FHMM~\cite{15RayNILM}, Seq2Point~\cite{Seq2Point}, and DAE~\cite{NeuralNILM}. 
These techniques are implemented in NILMTK~\cite{NILMTK}.
Following the recommendation of~\cite{NILM_resolution},
which explored the impact of the temporal resolution of data on the accuracy of NILM methods,
we use data with 1-minute resolution. 

We evaluate the performance of these 3 NILM methods in disaggregating the loads of 7 appliances, 
including the washing machine, microwave, air conditioner, furnace, fridge, dryer and dish washer, 
in each of the 6 homes.
Since a dryer is not present in 4 of these homes, 
we only report the results for the remaining 2 homes for this appliance.
We train appliance models using real home load and individual appliance load data 
collected between June 1 and June 22, 2018.
We then calculate the error of disaggregating each appliance's load in the test data (from June 23 to June 30, 2018)
with 5 different sets of input data,
including the true home load, the net load (i.e., home load - BTM solar generation), and 3~versions of the disaggregated home load obtained by applying our disaggregation method (1P+3SP), and the two baseline methods described earlier.

Figure~\ref{fig:NILM_average_error} shows the average RMSE for each appliance 
and the overall RMSE for all appliances in these homes.
Two important observations can be made.
First, solar disaggregation will improve the overall NILM accuracy 
but different appliances are affected to a different extent.
The overall RMSE for all appliances will be 0.446 if we directly apply Seq2Point 
(the best performing NILM method) on the net load data.
However, the RMSE will be 0.150 if we apply it to the disaggregated home load estimated by 1P+3SP, 
an impressive 66.2\% improvement in disaggregation accuracy.
We also observed 52.3\% and 22.0\% improvements for DAE and FHMM methods, respectively.
Among all appliances, the air conditioner shows the most significant improvement in accuracy, 
while the disaggregation accuracy of the washing machine, dryer, microwave, and dish washer 
does not exhibit a statistically significant change.
This may be due to the fact that these appliances are usually used after the sunset, 
when the net load is equal to the home load,
while the air conditioner is used throughout the day.
It is also worth mentioning that the fridge yields quite similar performance 
for all 5 different types of input data using the 3 NILM techniques.
Our hypothesis is that this is because the fridge has a distinct power consumption pattern 
that is easier to detect.

Our second observation is that a higher accuracy in solar disaggregation 
leads to a better NILM performance, 
especially for appliances with more variable power usage patterns (e.g., the air conditioner).
The overall RMSE of Seq2Point when it runs on the disaggregated home load 
obtained by the 1P+3SP method is 22.3\% and 9.7\% lower than 
when it runs on the disaggregated home load 
obtained by applying Baseline~1 and Baseline~2, respectively.
Among the 7 appliances, the NILM performance improvement achieved by our method over the two baselines 
is particularly noticeable for air conditioner,
which is the most power-hungry appliance in our study.

From the results depicted in Figure~\ref{fig:NILM_average_error} 
it is clear that Seq2Point yields a better performance than the other 2 NILM methods.
To further study the effect of solar disaggregation on the the performance of NILM, 
we compare the air conditioning load obtained by applying Seq2Point to the net load, true home load, and disaggregated home load using our 1P+3SP method in Figure~\ref{fig:NILM_appliance}(A).
We also show the true air conditioning load as a point of reference.
We can see that the error of directly applying Seq2Point to the net load data is specially high 
around noon which solar generation is usually at its peak.
Figure~\ref{fig:NILM_appliance}(B) compares the air conditioner load obtained by applying Seq2Point to 
the home load disaggregated by 1P+3SP and the 2 baselines.
It can be seen that among the three solar disaggregation methods,
our method enables tracking the true air conditioner's demand more closely.


\section{Conclusion}
Solar disaggregation will be essential for the reliable operation 
of a nimble and transparent power distribution system.
We proposed a method for disaggregating solar generation of BTM PV systems in an offline fashion.
This method relies on net load data and proxy measurements 
from a few solar PV systems located in the same geographic area as the target home.
We evaluated our proposed method on two different dataset from two countries with different climates.
We found that the solar estimation accuracy improves by 15.66\% on average over the best baseline (i.e., Baseline~2)
when one real proxy and three synthetic proxies are used.
Finally, we investigated whether a more accurate disaggregation technique 
could lead to higher accuracy in NILM.
Our results suggest that using the disaggregated home load rather than the net load 
improves the overall accuracy of NILM by 22.0\% to 66.2\%.

There are several avenues for future work.
First, we plan to test our method in a different climate 
where snow accumulation can greatly affect the output of PV systems.
We will explore how the proposed solar disaggregation technique can be used to 
estimate and predict the total amount of solar power generated in a neighbourhood.
Another interesting direction that will be pursued
is to design a solar disaggregation method that can be used in the presence of BTM battery storage.


\bibliographystyle{ACM-Reference-Format}
\bibliography{paper_ref}

\appendix

\end{document}